\documentclass{svproc}
\usepackage{url}
\usepackage[utf8x]{inputenc}
\usepackage{graphicx}

\begin{document}
\mainmatter              
\title{Object-Process Methodology for Intelligent System Development}
\titlerunning{OPM for Intelligent System Development}  
%
\author{V. P. Dorofeev}
\authorrunning{V. P. Dorofeev} 
%
\tocauthor{Vladislav Dorofeev}
\institute{Scientific Research Institute for System Analysis of Russian Academy of Sciences, Moscow, Russia,\\
\email{vladislav.p.dorofeev@yandex.ru}}

\maketitle              

\begin{abstract}
Development of the new artificial systems with unique characteristics is very challenging task. In this paper the application of the hybrid super intelligence concept with object-process methodology to develop unique high-performance computational systems is considered. The methodological approach how to design new intelligent components for existing high-performance computing development systems is proposed on the example of system requirements creation for "MicroAI" and "Artificial Electronic" systems.
\end{abstract}
\section{Introduction}
\begin{figure}
\center{\includegraphics[scale=0.5]{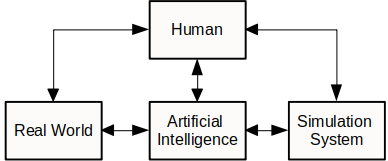}}
\caption{Hybrid super intelligent system}
\label{ris:sird}
\end{figure}

Hybrid intelligent systems are considered one of the most promising areas for the development of artificial intelligence \cite{stone:brook}. \cite{dor:leb}, \cite{dor:2021} introduced a hybrid super intelligence model for solving complex real-world problems. The common architecture of this model is shown on the Fig. \ref{ris:sird}. The presented model has strong empirical confirmation in research of complex natural systems, such as the fight against the COVID-19 pandemic \cite{alsd:beng} \cite{gupt:mah}, in the treatment of cardiac diseases \cite{bar:gas} and in the Systems Earth Science \cite{stef:rich}.

In this article will be considered the application of this architecture to creation of new artificial systems with the aid of systematic approach. The system science and systematic approach were born in the middle of the 20th century for the interdisciplinary research of complex natural and artificial systems \cite{bert:systh}, \cite{rand:sys}, \cite{hall:syseng}, \cite{booch:hist}. Today Model-based Systems Engineering \cite{mic:mbse} is the dominant approach in the area of creation the new artificial systems.

At the same time, there is a process of integration of methods for creating different types of artificial systems within the paradigms of Systems of Systems, Socio-technical systems, Cyberphysical systems, etc \cite{incose:2015}. Naturally, the question arises about the integration of modeling systems used in different areas. In the field of software, the leading standard is UML \cite{uml:link}, in the field of cyber-physical systems SysML \cite{sysml:link} is used. DoDAF \cite{dodaf:link}, TOGAF \cite{togaf:link} and others are used to design enterprise architectures. \cite{dori:book} proposes one such unifying method based on an object-process methodology. This method allows modeling the structure and behavior of systems with a minimal ontology (objects, states of objects, processes and connections between them are used) within the framework of a single type of model. At the same time, a modeling language such as SysML uses 9 types of models. The object-process method is now evolving and finds applications in a wide variety of fields \cite{dori:book}. A similar ontology has been used by SAP for creating enterprise applications (master data, business transactions) \cite{sap:help} and has proven to be effective.

This paper considers an example of using the object-process methodology to analyze and improve the development process of unique high-performance systems. Unique high-performance systems are understood as systems with unique technical characteristics, the development of which implies not only the use of existing methods and tools, but also their refinement, as well as the creation of new ones. An example of a unique high-performance system is a high-performance system based on the Cerebras Systems - 2 chip with 2.5 trillion transistors and a specially modified Tensorflow for it \cite{cerebras:systems}.

\section{Object-Process Methodology for System Engineering}

\begin{figure}
\center{\includegraphics[scale=0.5]{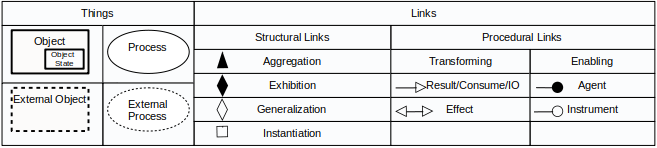}}
\caption{OPM objects}
\label{opm:pics}
\end{figure}

The object-process methodology \cite{dori:book} for describing systems is based on the idea of ​​minimizing the ontology required to describe the structure and behavior of a system. Its main components are: objects, states of objects, processes, links. Objects are intended to display the structure of the system, processes are intended to display the behavior of the system. Links can be structural (aggregation, exhibition, generalization, instantiation) and procedural (transformimng, enabling).
On the Fig. \ref{opm:pics} the LibreOffice Draw images that represent OPM objects are shown.

There are two types of system descriptions that are used together: diagrams and text descriptions. The description of the system in the object-process methodology allows to automatically translate the model into a graph representation format and simplifies its formal analysis \cite{ref:shan} with OPCloud solution.

In this work OPM diagrams are used only. In the next section the capabilities of OPM Modeling are shown on the example of the Unique HPC System Development.

\section {Object-Process Model of the Unique HPC System Development}
\begin{figure}
\center{\includegraphics[scale=0.5]{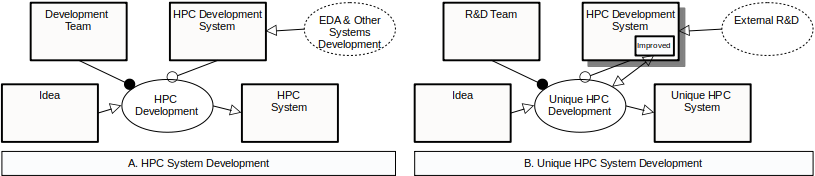}}
\caption{Comparison of HPC System Development and Unique HPC System Development}
\label{uopm:dev}
\end{figure}

The Unique High-Performance System is the system with superior characteristics to existing systems that is developed with the best technologies and scientific results.
The OPM diagrams on Fig. \ref{uopm:dev} are used to explain the distinction of High-Performance System and Unique High-Performance System. HPC Development System is constantly developing with new capabilities and the development team of this system is the part of Unique HPC System Development.

\begin{figure}
\center{\includegraphics[scale=0.6]{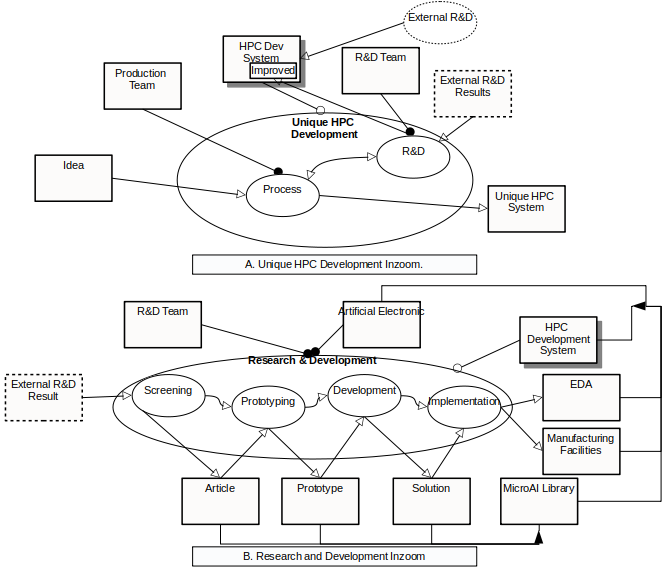}}
\caption{Unique HPC Development Process Detalization}
\label{hpc:inz1}
\end{figure}

At Fig. \ref{hpc:inz1}.A the process of Unique HPC system creation is in-zoomed by 1 level and divided on two processes: "Production" and "Research and Development". For development pf the Unique high-performance computational system the "Research and Development" process is very important. The main function of this process is the development of the most effective technologies and instruments for production on the base of results of the external results of research and development and own research.

At Fig. \ref{hpc:inz1}.B the process "Research and development" is unfolded at the next level. It contains following processes: "Screening", "Prototyping", "Development" and "Implementation". Here the enabling systems "MicroAI" and "Artificial Electronic" apear. The functions of the "MicroAI" system are: "Articles Storage", "Prototypes Storage" and "Solutions Storage". The functions of "Artificial Electronic" systems are: "Screening Assistance", "Prototyping Assistance", "Development Assistance" and "Implementation Assistance". Consequent detalization of this processes allows to make exact specification requirements for these systems.

Thus, the object-process methodology allows you to easily and efficiently obtain the system context and requirements specifications for the new systems.

\section{Conclusion}

Model-Based System Engineering in high-performance computing is the one of the most effective living examples of Hybrid Super Intelligence Systems. The results of the work of such intelligence are well measurable in terms of performance, energy consumption and the size of the systems being created, which gives it additional advantages in development. The hybrid intelligent system could be considered as the pragmatic way to super artificial intelligent system by continual automation of system engineering task.
This paper demonstrates a simple and effective approach for the development of intelligent information systems using object-process methodology. This approach is possible with the open source LibreOffice Draw toolkit. In combination with PostgreSQL, a full-fledged modeling system realization is possible.

\paragraph{Acknowledgements}
The work financially supported by State Program of SRISA RAS No. 0065-2019-0003 (AAA-A19-119011590090-2).

%
%


\begin{thebibliography}{6}
%

\bibitem {stone:brook}
Stone P., Brooks R., Brynjolfsson E., Calo R., Etzioni O., Hager G., and Leyton-Brown K. 2016. 
Artificial intelligence and life in 2030: One hundred year study on artificial intelligence: Report of the 2015–2016 Study Panel. 
\url {https://ai100.stanford.edu/2016-report}

\bibitem {dor:leb}
V. Dorofeev, A. Lebedev, V. Shakirov, W. Dunin-Barkowski. 2020. Super Intelligence to solve COVID-19 Problem. Advances in Neural Computation, Machine Learning, and Cognitive Research IV pp 293-300.

\bibitem {dor:2021}
V. Dorofeev. 2021. Hybrid Super Intelligence and Principles of Legasov, Efremov and Moiseev. (In Russian).

\bibitem {alsd:beng}
Hannah Alsdurf, Yoshua Bengio, Tristan Deleu, Prateek Gupta, Daphne Ippolito, Richard Janda, Max Jarvie, Tyler Kolody, Sekoul Krastev, Tegan Maharaj, Robert Obryk, Dan Pilat, Valerie Pisano, Benjamin Prud'homme, Meng Qu, Nasim Rahaman, Irina Rish, Jean-Franois Rousseau, Abhinav Sharma, Brooke Struck, Jian Tang, Martin Weiss, Yun William Yu. 2020. Covi white paper. ArXiv:2005.08502.

\bibitem {gupt:mah}
Prateek Gupta, Tegan Maharaj, Martin Weiss, Nasim Rahaman, Hannah Alsdurf, Abhinav Sharma, Nanor Minoyan, Soren Harnois-Leblanc, Victor Schmidt, Pierre-Luc St Charles, Tristan Deleu, Andrew Williams, Akshay Patel, Meng Qu, Olexa Bilaniuk, Gaétan Marceau Caron, Pierre Luc Carrier, Satya Ortiz-Gagné, Marc-Andre Rousseau, David Buckeridge, Joumana Ghosn, Yang Zhang, Bernhard Schölkopf, Jian Tang, Irina Rish, Christopher Pal, Joanna Merckx, Eilif B Muller, Yoshua Bengio. 2020. COVI-AgentSim: an Agent-based Model for Evaluating Methods of Digital Contact Tracing. ArXiv:2010.16004

\bibitem {bar:gas}
Barbara Rita Baricelli, Elena Gasiraghi, Daniela Fogli. 2019. A Survey on Digital Twin: Definitions, Characteristics, Applications, and Design Implications. IEEE Access. 
\url {DOI: 10.1109/ACCESS.2019.2953499}

\bibitem {stef:rich}
Will Steffen, Katherine Richardson, Johan Rockström, Hans-Joachim Schellnhuber, Opha Pauline Dube, Sébastien Dutreuil, Timothy M. Lenton, and Jane Lubchenco. 2020. The emergence and evolution of Earth System Science. Nature Reviews Earth and Environment. \url {http://hdl.handle.net/10871/40416}

\bibitem {bert:systh}
L. Bertalanfy. 1968. General System Theory.

\bibitem {rand:sys}
Malcolm W. Hoag. 1956. An Introduction to Systems Analysis. RAND Corporation.

\bibitem {hall:syseng}
A. Hall. 1965. A Methodology for System Engineering.

\bibitem {booch:hist}
G. Booch. 2018. The History of Software Engineering.
\url {DOI: 10.1109/MS.2018.3571234}

\bibitem {mic:mbse}
P. Micouin. 2014. Model-Based Systems Engineering. Fundamentals and Methods. Wiley.

\bibitem {incose:2015}
INCOSE. 2015. INCOSE Systems Engineering Handbook. 

\bibitem {uml:link}
Unified Modeling Language.
\url {https://www.uml.org/}

\bibitem {sysml:link}
System Modeling Language.
\url {https://sysml.org/}

\bibitem {dodaf:link}
The DoDAF Architecture Framework.
\url {https://dodcio.defense.gov/Library/DoD-Architecture-Framework/}

\bibitem {togaf:link}
The Open Group Architecture Framework.
\url {https://www.opengroup.org/togaf}

\bibitem {dori:book}
D. Dori. 2015.  Model-Based Systems Engineering with OPM and SysML. Springer.

\bibitem {sap:help}
SAP Help
\url {https://help.sap.com}

\bibitem {cerebras:systems}
Cerebras Systems.
\url {https://www.cerebras.net/product/}

\bibitem {ref:shan}
D. Medvedev, U. Shani and D. Dori. 2021. Gaining Insights into Conceptual Models: A Graph-Theoretic Querying Approach. Appl. Sci. 2021, 11, 765. 
\url {https://doi.org/10.3390/app11020765}


\end{thebibliography}
\end{document}